
\magnification=1200
\hsize=6truein\vsize=8.5truein

\font\open=msbm10 

\font\bigbf=cmbx10 scaled\magstep1


\def\mbox#1{{\leavevmode\hbox{#1}}}
\def\hspace#1{{\phantom{\mbox#1}}}
\def\oR{\mbox{\open\char82}}
\def\oZ{\mbox{\open\char90}}

\def\P{{\rm P}}

\def\O{{\rm O}}
\def\rS{{\rm S}}

\def\la{\lambda}

\def\zf{$\zeta$--function}
\def\zfs{$\zeta$--functions}

\def\frac#1/#2{\leavevmode\kern.1em
\raise.5ex\hbox{\the\scriptfont0 #1}\kern-.1em/\kern-.15em
\lower.25ex\hbox{\the\scriptfont0 #2}}
\def\sfrac#1/#2{\leavevmode\kern.1em
\raise.5ex\hbox{\the\scriptscriptfont0 #1}\kern-.1em/\kern-.15em
\lower.25ex\hbox{\the\scriptscriptfont0 #2}}

\def\gtorder{\mathrel{\raise.3ex\hbox{$>$}\mkern-14mu
             \lower0.6ex\hbox{$\sim$}}}
\def\ltorder{\mathrel{\raise.3ex\hbox{$<$}|mkern-14mu
             \lower0.6ex\hbox{\sim$}}}

\def\semidirprod{\rlap{\ss C}\raise1pt\hbox{$\mkern.75mu\times$}}

\def\for{\lower6pt\hbox{$\Big|$}}
\def\fish{\kern-.25em{\phantom{abcde}\over \phantom{abcde}}\kern-.25em}

\def\scdot{{\cdot}}
\def\boxit#1{\vbox{\hrule\hbox{\vrule\kern3pt
        \vbox{\kern3pt#1\kern3pt}\kern3pt\vrule}\hrule}}
\def\dalemb#1#2{{\vbox{\hrule height .#2pt
        \hbox{\vrule width.#2pt height#1pt \kern#1pt
                \vrule width.#2pt}
        \hrule height.#2pt}}}

\def\textsum{{\textstyle \sum}}


\def\noin{\noindent}

\def\al{\alpha}
\def\be{\beta}
\def\ga{\gamma}
\def\de{\delta}
\def\Ga{\Gamma}

\def\ka{\kappa}
\def\la{\lambda}

\def\th{\theta}
\def\ze{\zeta}

\def\comb#1#2{{\left(#1\atop#2\right)}}

\def\eg{{\it e.g. }}
\def\ie{{\it i.e. }}
\def\cf{{\it cf }}
\def\pa{\partial}

\def\gap{\vskip 20truept}

\def\tr{{\rm tr}}

\def\wR{{\widehat R}}

\def\sect{{\vskip 10truept\noindent}}

\def\3j#1#2#3#4#5#6{\left\lgroup\matrix{#1&#2&#3\cr#4&#5&#6\cr}
\right\rgroup}


\def\nolabels{\def\eqnlabel##1{}\def\eqlabel##1{}\def\reflabel##1{}}
\def\writelabels{\def\eqnlabel##1{%
{\escapechar=` \hfill\rlap{\hskip.09in\string##1}}}%
\def\eqlabel##1{{\escapechar=` \rlap{\hskip.09in\string##1}}}%
\def\reflabel##1{\noexpand\llap{\string\string\string##1\hskip.31in}}}
\nolabels
\global\newcount\meqno \global\meqno=1
\global\meqno=1
\def\eqnn#1{\xdef #1{(\the\meqno)}%
\global\advance\meqno by1\eqnlabel#1}
\def\eqna#1{\xdef #1##1{\hbox{$(\the\meqno##1)$}}%
\global\advance\meqno by1\eqnlabel{#1$\{\}$}}
\def\eqn#1#2{\xdef #1{(\the\meqno)}\global\advance\meqno by1%
$$#2\eqno#1\eqlabel#1$$}


\global\newcount\refno
\global\refno=1 \newwrite\reffile
\newwrite\refmac
\newlinechar=`\^^J
\def\ref#1#2{\the\refno\nref#1{#2}}
\def\nref#1#2{\xdef#1{{\bf\the\refno}} 
\ifnum\refno=1\immediate\openout\reffile=refs.tmp\fi
\immediate\write\reffile{
     \noexpand\item{[{\noexpand#1}]\ }#2\noexpand\nobreak.}
     \immediate\write\refmac{\def\noexpand#1{\the\refno}}
   \global\advance\refno by1}
\def\semi{;\hfil\noexpand\break ^^J}
\def\refn#1#2{\nref#1{#2}}

\def\cmp#1#2#3{{\it Comm. Math. Phys.} {\bf {#1}} (19{#2}) #3}

\def\np#1#2#3{{\it Nucl. Phys.} {\bf B{#1}} (19{#2}) #3}

\def\prA#1#2#3{{\it Phys. Rev.} {\bf A{#1}} (19{#2}) #3}

\def\prD#1#2#3{{\it Phys. Rev.} {\bf D{#1}} (19{#2}) #3}

\def\zfn#1#2#3{{\it Z.f.Naturf.} {\bf {#1}} (19{#2}) #3}

\def\asens#1#2#3{{\it Ann. Sci. \'Ecole Norm. Sup. (Paris)} {\bf{#1}} (#2) #3}

\def\pcps#1#2#3{{\it Proc. Camb. Phil. Soc.} {\bf{#1}} (19{#2}) #3}

\def\amsh#1#2#3{{\it Abh. Math. Sem. Ham.} {\bf {#1}} (19{#2}) #3}
\def\amm#1#2#3{{\it Am. Math. Mon.} {\bf {#1}} (19{#2}) #3}

\def\cjm#1#2#3{{\it Can. J. Math.} {\bf {#1}} (19{#2}) #3}
\def\dmj#1#2#3{{\it Duke Math. J.} {\bf {#1}} (19{#2}) #3}
\def\jdg#1#2#3{{\it J. Diff. Geom.} {\bf {#1}} (19{#2}) #3}
\def\ma#1#2#3{{\it Math. Ann.} {\bf {#1}} ({#2}) #3}

\def\plms#1#2#3{{\it Proc. Lond. Math. Soc.} {\bf {#1}} (19{#2}) #3}

\def\qjm#1#2#3{{\it Quart. J. Math.} {\bf {#1}} (19{#2}) #3}
\def\qjpam#1#2#3{{\it Quart. J. Pure and Appl. Math.} {\bf {#1}} ({#2}) #3}
\def\rcmp#1#2#3{{\it Rend. Circ. Mat. Palermo} {\bf {#1}} (19{#2}) #3}

\def\tams#1#2#3{{\it Trans. Am. Math. Soc.} {\bf {#1}} (19{#2}) #3}
\refn\Dowk{J.S.Dowker {\it Effective action in spherical domains} MUTP/93/15}
\refn\Cesaroa{G.Cesaro {\it Mineralogical Mag.} {\bf 17} (1915) 173}
\refn\Fedorova{E.S.Fedorov {\it Mineralogical Mag.} {\bf 18} (1919) 99}
\refn\Fedorovb{E.S.Fedorov {\it Symmetry of Crystals} Translated by D. and K.
Harker, American Crystallographic Association (1971)}
\refn\Cesarob{G.Cesaro {\it `Des macles'. M\'emoires couronn\'es et
m\'emoires des savants
\'etrangers publi\'es par l'Acad\'emie royale de Belgique} {\bf 53} (1893) 47}
\refn\Coxetera{H.S.M.Coxeter {\it Regular Polytopes} Methuen, London (1948)}
\refn\Klein{F.Klein {\it Lectures on the icosahedron} Methuen, London (1913)}
\refn\Steinberg{R.Steinberg \cjm {10}{58}{220}}
\refn\Coxeteraa{H.S.M.Coxeter {\it Regular Polytopes} 2nd ed. MacMillan,
New York (1963)}
\refn\Coxeterg{H.S.M.Coxeter \dmj {18} {51} {765}}
\refn\Coxeterc{H.S.M.Coxeter and W.O.J.Moser {\it Generators and
relations for finite groups} Springer-Verlag, Berlin (1957)}
\refn\Chang{Peter Chang and J.S.Dowker \np {395}{93}{407}}
\refn\Shephard{G.C.Shephard and J.A.Todd \cjm{6}{54}{274}}
\refn\Solomon{L.Solomon {\it Nagoya Math. J.} {\bf 22} (1963) 57}
\refn\Barnes{E.W.Barnes {\it Trans. Camb. Phil. Soc.} (1903) 376}
\refn\Bateman{A.Erdelyi,W.Magnus,F.Oberhettinger and F.G.Tricomi {\it Higher
Transcendental functions} McGraw-Hill, New York (1953)}
\refn\Brownell{F.H.Brownell {\it J.Math.Mech.} {\bf 6} (1957) 119}
\refn\Fedosov{B.V.Fedosov {\it Sov. Mat. Dokl.} {\bf 4} (1963) 1092; {\it ibid}
{\bf 5} (1964) 988}
\refn\Kac{M.Kac \amm {73}{66}1}

\refn\Laporte{O.Laporte \zfn {\bf 3a} {48} 447}
\refn\Sommerville{D.M.Y.Sommerville {\it Geometry of n-dimensions} Methuen,
London (1929)}
\refn\Coxeterb{H.S.M.Coxeter {\it Regular Complex Polytopes} 2nd. Edn.
Cambridge University Press, Cambridge (1991)}
\refn\Berger{M.Berger {\it Geometry} vol. II, Springer, Berlin (1989)}
\refn\Witt{E.Witt \amsh {14}{41}{289}}
\refn\Goursat{E.Goursat \asens {16}{1889}{9}}
\refn\Steinbergb{R.Steinberg \tams {103}{58}{493}}
\refn\Todd{J.A.Todd \pcps {27}{30}{212}}
\refn\Sommervilleb{D.M.Y.Sommerville \rcmp {48}{24}9}
\refn\Grunbaum{B.Gr\"unbaum {\it Convex Polytopes} Interscience, New York,
(1967)}
\refn\Coxetere{H.S.M.Coxeter \pcps{27}{30}{201}}
\refn\Schl{L.Schl\"afli \qjpam {2}{1858}{217}; {\it ibid} {\bf 3} (1860) 54,
97}
\refn\Richmond{H.W.Richmond \qjpam{34}{1903}{175}}
\refn\Coxeterf{H.S.M.Coxeter \qjm{6}{35}{417}}
\refn\Schlb{L.Schl\"afli {\it Gesammelte Mathematische Abhandlungen} vol.I.
Birkh\"auser, Basel (1950)}
\refn\Leese{R.A.Leese and T.Propopec \prD{44}{91}{3749}}
\refn\Laursen{M.L.Laursen, G.Schierholz and U.-J.Wiese \cmp{103}{86}{693}}
\refn\Woit{P.Woit \np{262}{85}{284}}
\refn\Pathria{R.K.Pathria {\it Suppl.Nuovo Cim.} {\bf 4} (1966) 276}
\refn\Baltes{H.P.Baltes \prA{6}{72}{2252}}
\refn\Cheeger{J.Cheeger \jdg {18}{83}{575}}

\refn\Rub{A.Rubinowicz \ma {96}{1926}{648}}
\refn\Hirz {F.Hirzebruch{\it Topological methods in algebraic geometry},
3rd. edition, Springer-Verlag, Berlin (1978)}
\refn\Toddb{J.A.Todd \plms {43}{37}{190}}


\vglue 1truein
\rightline {MUTP/93/20}
\gap
\centerline {\bigbf A zeta function approach to the relation}
\centerline{\bigbf  between the numbers of symmetry}
\centerline {\bigbf planes and axes of a polytope}
\vskip 15truept
\centerline{J.S.Dowker}
\vskip 10 truept
\centerline {Department of Theoretical Physics,}
\centerline{The University of Manchester, Manchester, England.}
\vskip 40truept
\centerline {Abstract}
\vskip 10truept
A derivation of the Ces\`aro-Fedorov relation from the Selberg trace formula on
an orbifolded 2-sphere is elaborated and extended to higher dimensions using
the
known heat-kernel coefficients for manifolds with piecewise-linear boundaries.
Several results are obtained that relate the coefficients, $b_i$, in the
Shephard-Todd polynomial to the geometry of the fundamental domain. For the
3-sphere we show that $b_4$ is given by the ratio of the volume of the
fundamental tetrahedron to its Schl\"afli reciprocal.
\vskip 15truept
\rightline {September 1993}
\vfill\eject
\noin{\bf 1. Introduction}.

\noin In an earlier work [\Dowk] we have indicated how the relation between the
numbers of symmetry planes and symmetry axes of a solid discovered by Ces\`aro
[\Cesaroa], and Fedorov [\Fedorova,\Fedorovb] can be derived analytically.
In the present work we wish to further analyse the method and extend it to
higher dimensions.

This work can be looked upon as a discussion of the heat-kernel and \zf\ on
certain manifolds and so should be relevant to quantum theory.

Ces\`aro's proof
of the formula relies on an earlier result of his [\Cesarob] relating the
number of symmetry planes to the number and order of symmetry axes lying in a
single symmetry plane. For completeness we repeat his findings here. Let $b_1$
be the total number of symmetry planes and let there be, on the $i$-th such
plane, $n^{(i)}_m$ symmetry axes of order $m$, then Ces\`aro [\Cesarob]
shows that (see later)
\eqn\ces{
b_1=1+\sum_m n^{(i)}_m(m-1),\quad \forall i.
}
Summing this formula over all symmetry planes and using the fact that
$$\sum_in^{(i)}_m=mn_m,$$
where $n_m$ is the total number of symmetry axes of order $m$, we find one form
of the C-F relation,
\eqn\CFa{
b_1(b_1-1)=\sum_mn_mm(m-1).
}
Fedorov [\Fedorova,\Fedorovb] refers, somewhat confusingly, to $x(x-1)$ as
the `function' of $x$. \CFa\ is his `zonohedral formula'. (See [\Coxetera].)


Another form of the relation follows by noting that, for the regular
solids at least, we have the orbit-stabiliser result
\eqn\reg{
g=2mn_m,\quad\forall m,
}
where $2g$ is the order of the complete symmetry group, $\Ga$.
Therefore
\eqn\CFb{
2b_1(b_1-1)=g\sum_m(m-1)=g(p+q+r-3)
}
for the three types of axis of orders $p$, $q$ and $r$.

Relations \CFa\ and \CFb, and their generalisations, are our interest in the
present paper. Because the theory of polytopes might not be too
familiar, a certain amount of historical and technical commentary might
be of benefit, if only to the writer. Everything goes back to Schl\"afli.
\vfill\eject
\sect{\bf 2. A survey of the geometrical method}

\noin We recall the standard geometric situation [\Klein], [\Coxetera]
\S$3\scdot5,$ \S$4\scdot5$. 
The regular polyhedron $\{p,q\}$ is projected onto its circumsphere
giving a spherical tessellation. The planes of symmetry of $\{p,\,q\}$
divide this tessellation further into $|\Ga|$ congruent spherical (M\"obius)
triangles, $(pqr)$ (with interior angles $\pi/p$, $\pi/q$ and $\pi/r$),
which are transitively permuted under the action of
$\Ga$, $=[p,\,q]$. This produces a simplicial
decomposition (triangulation) of the sphere. The `$q$-vertex' ({\bf 0}) of
$(pqr)$,
and its images under $\Ga$, correspond to the vertices of $\{p,q\}$ and the
`$p$-vertex' ({\bf 2}) likewise to the vertices of the reciprocal solid,
$\{q,\,p\}$, \ie to the
mid-points of the faces of $\{p,\,q\}$. The `$r$-vertex' ({\bf 1}) is
associated
with the mid-points of the edges and, for the usual solids, $r=2$ so that
$g=4n_r$.  The edge mid-points form the vertices of the quasi-regular
polyhedron $\left\{{\textstyle p}\atop {\textstyle q}\right\}$ which
projects to a spherical tessellation of edge length $2\pi/h$, where $h$
is the Coxeter number ([\Coxetera] \S$2\scdot3$, \S$2\scdot6,\,5\scdot91$).

Equation \reg\ says that there are $g/m$ images of the $m$-fold vertex of
$(pqr)$ under $[p,\,q]$. This follows because the stability group of this
vertex has order $2m$.

Steinberg [\Steinberg] gives a simple geometrical derivation of \CFa\
which appears in the second edition of Coxeter's book [\Coxeteraa].
In the first edition, [\Coxetera], Coxeter just indicates the general
nature of
such a proof. Curiously, in [\Coxeteraa], he drops the references to
Ces\`aro and Fedorov, presumably because their relations are superseded by
Steinberg's and some further work of his own, [\Coxeterg].

The geometrical argument is worth repeating.
The $b_1$ symmetry circles intersect in $b_1(b_1-1)$ (possibly coincident)
points (the vertices of the triangulation).
In order that a vertex correspond to a symmetry axis of order $m$, it is
necessary that $m$ symmetry circles cross there. That is to say,
there must be a coincidence of $m(m-1)/2$ intersections at the vertex.
Equation \CFa\ expresses the distribution of the total number of intersections
among such vertices, and there are two vertices per axis. We might term
$m(m-1)/2$ the {\it degeneracy} of the intersection.

It is clear that this argument is essentially the same as Ces\`aro's.
Indeed
\ces\ follows from a counting argument applied to a single reflecting circle.
The total number of such circles is just one more than the
number that
cross a chosen circle. Further, trivially, the number that cross this circle
at a vertex of order $m$ is $m-1$ (discounting the original one).

We might note the
relations between the $n_m$ and the numbers of vertices, edges, and faces,
$N_0$, $N_1$ and $N_2$ respectively, of the polyhedron $\{p,q\}$,
([\Coxetera] $5\scdot81$),
\eqn\vef{
2n_p=N_0,\qquad2n_q=N_2,\qquad2n_r=N_1.
}
The statement that $qN_0=2N_1=pN_2$ is a graph-theoretic one (see \eg
[\Coxeterc] Chap.8).

It is worth outlining the further development for
completeness [\Coxetera], [\Steinberg]. The associated polyhedron
$\left\{{\textstyle p}\atop {\textstyle q}\right\}$ has $N_1$
vertices, $2N_1$ edges (from graph theory) and $N_0+N_2$ faces
(of two sorts, $\{q\}$ and $\{p\}$).
Every edge belongs to just one equatorial $h$-gon of which, therefore, there
are $2N_1/h$ in total. Since there are four faces at each vertex, the
intersections of the $h$-gons are simple (degeneracy of one) and the
total number of intersections is just the total number of vertices, \ie
$${2N_1\over h}\big({2N_1\over h}-1\big)=N_1$$
Equivalently, the number of equators is one plus the number of intersections
with one equator \ie $2N_1/h=1+h/2$.
Thus
\eqn\order{
2g=h(h+2).
}

It is easily seen that each equator on $\{p,q\}$ contains $h$ points ${\bf 1}$
and crosses $h$ segments ${\bf 02}$.
The total number of reflecting circles on $\{p,q\}$ may now be obtained by
adding up the crossings on a complete circumnavigation of an equator. There
are two for each {\bf 1} point and one for each {\bf 02} segment crossed.
Taking into account the antipodal points we get $b_1=(2h+h)/2$,
\eqn\nocirc{
b_1={3h\over2}.
}

{}From \nocirc, \order\ and \CFb\ we find Steinberg's formula [\Steinberg],
[\Coxeteraa],
\eqn\stein{
h+2={24\over12-p-q-r}
}
for the Coxeter number in terms of the angles of the M\"obius triangle.
\sect{\bf 3. The analytical method}

\noin Let $\ze(s)$ be the \zf\ for the Laplacian on the orbifolded 2-sphere,
\ie on the M\"obius triangle $(pqr)$. The analytical method consists
of identifying $\ze(0)$ calculated in two ways -- one from the expression in
terms of the degrees {\bf d} of the symmetry group, $\Ga$, (this involves the
Barnes \zf) and another that comes {\it either} by using the expression for
the
constant term in the heat-kernel, derived by differential geometric methods for
{\it any} manifold and {\it then} applied to the orbifolded sphere, {\it or} by
using the particular
form of the \zf\ on the orbifolded sphere in terms of the symmetry axes,
\ie fixed points, as given in [\Chang].
We could say
that the Ces\`aro-Fedorov (C-F) formula is an application of the
Selberg trace formula to these spherical domains.

The formula referred to in [\Chang] gives a constructional form to the
\zf\ in terms of `cyclic' \zfs. As derived, the formula is for
the rotational subgroup $\Ga^+$ and a special conformal coupling. It comes
originally from the conjugacy class decomposition of the heat-kernel
and is actually independent of the coupling.

Although convenient, using this particular form of
$\ze$ on S$^2/\Ga$ is, in some
ways, not as satisfying as an appeal to a more universal
formula such as that for the constant term in the heat-kernel expansion used
in [\Dowk]. We need no knowledge of group theory nor
of conjugacy class decomposition to use or derive this formula. Of course, the
significance of the corner terms (see later) is, at root, a fixed point one.

The number of reflecting planes is introduced analytically through the
results of
Shephard and Todd [\Shephard] and Solomon [\Solomon] who give this
number as the sum of the {\it exponents}, $(d_i-1)$, of the  finite
reflection group. Our approach should thus be valid for solids other than
the regular ones.

Clearly the C-F relation would
emerge whatever the coupling (we shall verify this shortly) and also
there is no need to use the full, reflecting $\Ga$. We could have
obtained the C-F formula from the results of [\Chang].

Another point we shall investigate concerns the value of the \zf\ at other
arguments, in particular at negative integers. One would expect to get
different relations between the orders $m$ and the number of symmetry planes.
\sect{\bf 4. Conformal--rotational derivation}

\noin In this section we rederive the C-F relation using  the special
conformal coupling $\xi=1/8$  and the corresponding rotational
\zf\ on S$^2/\Ga$ derived in [\Chang].
We refer to equation (26) of [\Chang] which we give again with a slightly
different notation
\eqn\basic{
\ze(s)={1\over g}\left[\sum_mmn_m\ze_m(s)-2\big(\sum_mn_m-1\big)
\ze_R\big(2s-1,{1\over2}\big)\right].
}

$\ze_m(s)$ is the \zf\ for the rotational cyclic group, $\oZ_m$, and its
values at the
negative integers are shown in equation (32) of [\Chang]. (See the appendix
to the present paper). In particular we calculate
$\ze_m(0)$
\eqn\cyczero{
\ze_m(0)={2m^2-1\over12m}
}
and so we get from \basic,
\eqn\one{
\ze(0)=
{1\over12 g}\left[2\sum_mn_m(m^2-1)+1\right].
}

On the other hand we know that the rotational $\ze(s)$ is given by the
sum of the Neumann and Dirichlet \zfs,
\eqn\basicb{
\ze(s)=\ze_N(s)+\ze_D(s)
}
where
\eqn\zedefsn{
\ze_N(s)=\ze_d\big(2s,{d-1\over2}\mid {\bf d}\big)
}
\eqn\zedefsd{
\ze_D(s)=\ze_d\big(2s,\textsum d_i-{d-1\over2}\mid {\bf d}\big).
}
For $s$ a negative integer, or zero, the two terms in \basicb\ are equal.

Barnes [\Barnes] gives the values, (we use both Barnes' notation and the more
standard one of Erdelyi, [\Bateman], for the generalised Bernoulli
polynomials)
\eqn\barzero{
\ze_d(-k,a\mid{\bf d})={(-1)^d\over k+1}\!\!\!\!
\phantom{H}_dS_{1+k}^{(1)}(a)=
{(-1)^d\over\prod d_i}{k!\over(d+k)!}B^{(d)}_{d+k}(a\mid{\bf d})
}
where
$$\phantom{H}_dS_1^{(d+1)}(a)={1\over\prod d_i},\quad
\phantom{H}_dS_1^{(d)}(a)=
-{}_dS_1^{(d)}\big(\textsum d_i-a\big)=
{2a-\sum d_i\over2\prod d_i},$$

\eqn\poleres{
\phantom{H}_dS_1^{(d-1)}(a)=
{}_dS_1^{(d-1)}\big(\textsum d_i-a\big)=
{1\over12\prod d_i}\big(6a^2-6a\sum d_i+
\sum d_i^2+3\sum_{i<j}d_id_j\big)
}
and so we get from \basicb, \barzero\ and \poleres, with $g=d_1d_2$ and
$b_1=d_1+d_2-1$,
\eqn\two{
\ze(0)={1\over
12g}\left[3-6(d_1+d_2)+2\big(d_1^2+d_2^2+3d_1d_2\big)
\right]={1\over12g}\left[2b_1(b_1-1)-1+2g\right].
}
The equation that must hold, therefore, is
\eqn\reln{
b_1(b_1-1)=\sum_mn_m(m^2-1)+1-g.
}
Recalling the total rotational order equation
\eqn\totord{
\sum_m(m-1)n_m=g-1
}
we get
$$
b_1(b_1-1)=\sum_mn_mm(m-1)
$$
as required.

Compared with Steinberg's geometrical argument, this is a rather complicated
derivation. Actually, it is not so very different in content. We are simply
re-extracting the information that has already gone into the construction of
the \zf\ \basic. This remark leads onto the next point.
We note that the C-F relation has been derived in its first form, \CFa.
The derivation in [\Dowk] yields the second form \CFb. Let us see why
this is so.
\vfill\eject
\sect{\bf 5. Heat kernel method}

\noin For the differential operator $-\Delta^2+\xi R$,
the general formula for $\ze(0)$ on a two-dimensional domain, $\cal M$,
with boundary $\pa{\cal M}=\cup\,\pa{\cal M}_i$ is
\eqn\beeone{
\ze(0)={1-6\xi\over24\pi}\int_{\cal M} R\,dA+{1\over12\pi}\sum_i\int_
{\pa{\cal M}_i}\ka(l)\,dl+{1\over24\pi}\sum_{\th}{\pi^2-\th^2\over\th}
}
where the $\th$ sum runs over all inward facing angles at the corners
of $\pa\cal M$. For minimal coupling $\xi=0$, while for the special
conformal coupling in [\Chang], $\xi=1/8$.

The corner term was derived by Brownell [\Brownell] and Fedosov [\Fedosov]
and again by Kac [\Kac] as the constant term in the short-time expansion
of the heat-kernel. The angles $\th$ are quite general.

Restricting attention to the spherical triangle, ${\cal M}=(pqr)$,
$\ka$ must vanish since the sides are geodesic. Further,
if the triangle tiles the two-sphere once,
$${1-6\xi\over24\pi}\int_{\cal M} R\,dA={1-6\xi\over6g},$$
$2g$ being the number of triangles.

Choosing $\xi=1/8$ (not essential), so that $\ze(s)=\ze_N(s)$, the
identity becomes
\eqn\identtwoa{
g\sum_\th\left({\pi\over\th}-{\th\over\pi}\right)=2\big(b_1(b_1-1)+g-1\big)
}
or, writing the angle $\th$ typically as $\pi/m$,
\eqn\identtwob{
g\sum_m\left(m-{1\over m}\right)=2\big(b_1(b_1-1)+g-1\big).
}
Reorganising the left-hand side slightly just to see how things work out in
detail
\eqn\detail{
g\sum_m(m-1)+3g-g\sum_m {1\over m}=2\big(b_1(b_1-1)+g-1\big).
}
The Gauss-Bonnet, or area, equation reads
\eqn\euler{
\sum_m{1\over m}={2\over g}+1
}
and we see that the $2(g-1)$ term on the right of \detail\ will cancel against
a term on the left to regain \CFb, bypassing explicit mention of
the numbers $n_m$. We do have to know the area formula for a spherical
triangle, however.

The $n_m$ can be reinstated using \reg, when we see that \totord\ is
equivalent to \euler.

For future reference, we note that the quantity $g-1$ is the number of
elements of the complete symmetry group that fix an axis but not a plane
in $\oR^3$. It is just the number of nontrivial rotational elements \ie all
the elements in $\Ga^+$ except the identity.
We denote it by $b_2$ and write \totord\ as
\eqn\trivident{
b_2=\sum_mn_m(m-1)
}
which can be looked upon as an identity similar to, but more elementary
than, \CFa.

Further numerical expressions, which are easily obtained, give the
numbers of vertices, $V$, edges, $E$, and faces, $F$, of the triangulation
in terms of the group order, (\cf [\Coxetera] p 67)
\eqn\VFE{
V=2+g,\quad E=3g,\quad F=2g.
}

Equation \euler\ can be interpreted in the following way. Think of the
combination $2g\sum(1/2m)$ as a sum over all $2g$ fundamental domains of
a quantity, $\sum(1/2m)$, associated with one domain.
If $2g$ M\"obius triangles are assembled to form
the simplicial decomposition of $\rS^2$, each vertex will appear with a
multiplicity of $2m$. The $1/2m$ factor reduces this multiplicity to unity
and so $2g\sum(1/2m)$ equals the number of vertices, $V$.

\sect{\bf 6. Other relations. Higher dimensions}

\noin Other relations can be derived using \basicb\
and our general knowledge of the heat-kernel coefficients, or the
specific form of the \zf\ on S$^d/\Ga$, such as \basic.

The standard facts we need are that, if the heat-kernel expansion is

\eqn\hkern{
K(\tau)\cong{1\over(4\pi\tau)^{d/2}}\sum_{k=0,1/2,\ldots}^
\infty C_k\tau^k,
}
then
\eqn\negint{
\zeta(-k)={(-1)^kk!\over(4\pi)^{d/2}}C_{k+d/2}-n_0\de_{k0},
}
where $n_0$ is the number of zero modes, and, further, that
$\zeta_d(s)$ has poles at
$s=(d-m)/2$, for $m=0,1,\ldots,d-1$ and $m=d+1,d+3,\ldots$, with residues
\eqn\residues{
{(4\pi)^{-d/2}\over\Gamma\big((d-m)/2\big)}C_{m/2}.
}

These equations can be read in the two obvious ways -- knowing the values of
the \zf\ we can derive the coefficients and {\it vice versa}. For the moment we
adopt the latter viewpoint and use the expressions for the coefficients
developed using standard analytical methods. Our discussion is neither
comprehensive nor systematic.

Two basic dimension- and coupling-independent results are
\eqn\hkcoeffs{
C_0=|{\cal M}|,\quad C_{1/2}=
\pm{\sqrt\pi\over2}\sum_i|\pa{\cal M}_i|.
}
The $\pm$ signs are for Neumann and Dirichlet conditions.

{}From \residues\ we see that $C_0$ and $C_{1/2}$ are given by the residues
in $\ze(s)$ at $s=d/2$ and $s=(d-1)/2$ respectively.
For simplicity we might as well use the conformal \zfs\ given in
\basicb. Therefore we need the pole structure of the Barnes \zf\
[\Barnes] which is, as $s\rightarrow k/2$,
\eqn\bpoles{
\ze_d(2s,a\mid{\bf d})\rightarrow
{(-1)^{d+k}\over2(k-1)!}{\phantom{H}_dS_1^{(k+1)}(a)\over s-k/2}
}
where $k$ is one of $1,\,2,\,3,\,\ldots, d$.

Comparing \bpoles\ and \residues, \cf [\Chang] eqn. (76),
$$ C_{m/2}=(-1)^m{2^m\pi^{(d+1)/2}\over\Ga\big((d-m+1)/2\big)}\,
{}_dS^{(d-m+1)}_1(a)$$
\eqn\hkcoeff{
=(-1)^m2^{m-1}\pi^{m/2}|\rS^{d-m}|\,{}_dS^{(d-m+1)}_1(a)\,,
}
where $a$ is $(d-1)/2$ for Neumann conditions and $\sum d_i -(d-1)/2$ for
Dirichlet. For simplicity, we usually work with Neumann conditions.
In view of \poleres\ and \hkcoeffs\ we get in two dimensions, for example,
\eqn\firsttwo{
|{\cal M}|={2\pi\over d_1d_2},
\quad |\pa{\cal M}|
=2\pi{d_1+d_2-1\over d_1d_2}={2\pi b_1\over g}.
}

These equalities are easily confirmed from the trivial geometrical
fact that the
two-sphere is covered by $2g$ spherical triangles, the vertices of which are
formed by the intersections of the $b_1$ symmetry great circles. The sum of
the circumferences of these circles is half the total perimeter of
$2g$ fundamental domains, as is quickly appreciated by drawing in the sides
of each triangle separately -- equivalent to splitting each great circle
into two (\cf [\Coxetera] p232). More formally, one can use the
measure-theoretic principle
\eqn\measure{
|A\cup B|+|A\cap B|=|A|+|B|
}
applied to the edges.

Dividing in \firsttwo\ we find
$$b_1={|\pa{\cal M}|\over|{\cal M}|},$$
which is Laporte's rule, [\Laporte], for the number of symmetry circles, \ie
the nodal lines of the lowest odd mode (the Jacobian of degree $b_1$).

The extension of these elementary results to higher
dimensions is immediate. The relevant analytical fact is that, for all
$d$,
\eqn\relfact{
{}_dS^{(d+1)}_1(a)={1\over g},\quad {}_dS^{(d)}_1\big(
\textsum d_i-(d-1)/2\big)={b_1\over2g}
}
so that
\eqn\emvol{
{}|{\cal M}|={1\over2g}|\rS^d|
}
and
\eqn\deemvol{
{}|\pa{\cal M}|={b_1\over g}|\rS^{d-1}|.
}
The generalised Laporte rule is then
\eqn\Lapd{
b_1={|\rS^d|\over2|S^{d-1}|}{|\pa{\cal M}|\over|{\cal M}|}.
}

The geometric situation in $\oR^{d+1}$ is the generalisation of that in
$\oR^3$, recalled in section 1. We outline it here for any $d$
(see [\Sommerville],
[\Coxetera] pp 130, 137-140, [\Coxeterb] \S$4\scdot1$, [\Berger]
chap.12). We are most interested in $d=3$.

The projection of the $(d+1)$-dimensional regular
polytope, $\{k_1,\,k_2,\ldots,k_d\}$ (in $\oR^{d+1}$) onto its circumscribing
hypersphere yields a spherical tessellation, or honeycomb, the cells of which
are the projections of the $d$-faces of the polytope. The symmetry
$d$-flats (`reflecting hyperplanes') of $\{k_1,\,k_2,\,\ldots,k_d\}$
intersect the circumsphere, $\rS^d$, in
a set of  reflecting great $(d-1)$-spheres constituting the boundaries
of $|\Ga|$ spherical $d$-simplexes that are
transitively permuted by the complete symmetry group $\Ga,\,=[k_1,\,k_2,
\ldots,k_d]$ and which make up a simplicial subdivision of $\rS^d$, [\Witt].
(This decomposition could also be termed an irregular tessellation.)
The $(d-1)$-dimensional boundary of the fundamental domain
is the union of $d$ pieces of the reflecting great $(d-1)$-spheres.
Topologically, the honeycomb is the boundary, or frontier, of the `solid'
polytope.
For $d=3$ the fundamental domain is a spherical tetrahedron. Goursat [\Goursat]
gives a detailed account of the different $\rS^3$ honeycombs.

Note that the symbol $\{k_1,\,k_2,\ldots k_d\}$ can refer
to a $(d+1)$-dimensional polytope in $\oR^{d+1}$ or to a spherical
$(d+1)$-dimensional polytope in $\rS^{d+1}$ or to a $d$-dimensional honeycomb
on $\rS^d$, depending on context ([\Coxetera] p 138).

Equation \deemvol\ expresses the fact that the measure of {\it all} the
reflecting great spheres is one half the measure of the total boundary of
$2g$ fundamental domains.

The next identity arises on setting $m=2$ in  \hkcoeff,
\eqn\identone{
C_1=2\pi|\rS^{d-2}|\,{}_dS^{(d-1)}_1\big((d-1)/2
\big).
}

The $C_1$ coefficient in two dimensions has already been discussed in
section 2 but we wish to extend the argument to $d$-dimensions and so write
out its general form, valid for any coupling and region ${\cal M}$,
\eqn\ceeone{
C_1=
{1-6\xi\over6}\int_{\cal M} R\,dV+{1\over3}\sum_i\int_
{\pa{\cal M}_i}\ka\,dS+{1\over6}\sum_{i<j}
{\pi^2-\th_{ij}^2\over\th_{ij}}|\pa{\cal M}_i\cap\pa{\cal M}_j|.
}
Here $\th_{ij}$ is the (arbitrary) dihedral angle between the boundary
components $\pa{\cal M}_i$ and $\pa{\cal M}_j$. For simplicity we have
assumed that this angle is constant along the intersection. This
is so for a fundamental domain. Fedosov's calculation, [\Fedosov], of the final
term is for a flat polyhedral region so that the $\th_{ij}$ are constant for
him. For $d=3$ the term could be referred to as an edge term.

We again employ the conformal coupling of [\Chang] so that $\xi=(d-1)/4d$.
For ${\cal M}$ a fundamental domain, the first term simplifies to
\eqn\firstt{
 {1-6\xi\over6}\int_{\cal M} R\,dV={(3-d)(d-1)\over24g}|\rS^d|
 }
and, because the boundary is geodesic, the second term in \ceeone\ vanishes.

We now look at the right-hand side of \identone\ which can be written
more geometrically in terms of the number of reflecting $d$-planes  and
the number of those group elements that fix a $(d-1)$-plane but {\it not} a
$d$-plane. We denote these numbers by $b_1$ and $b_2$ respectively.
A reflecting $(d-1)$-plane intersects the circumsphere in a
reflecting great $(d-2)$-sphere.

The complete symmetry group $\Ga$ is generated by $(d+1)$ reflections,
$R_1,\ldots,$ $R_{d+1}$, in $(d+1)$ hyperplanes in $\oR^{d+1}$.
Generally, $b_r$ is the number of elements of $\Ga$
that fix a $(d+1-r)$-flat, but no flat of higher dimension,
in $\oR^{d+1}$. Equivalently, $b_r$ is the number of elements that are
expressible as products of $r$ (but no fewer) reflections, \ie  those that
have length $r$.

The Coxeter element,
$R_1R_2\ldots R_{d+1}$, has period $h$, the Coxeter number, and
characteristic roots $\exp(2\pi im_j/h)$ where the $m_j$ $(j=1,\,\ldots,
d+1)$ are the exponents, related to the degrees by $m_j=d_j-1$.


Solomon's theorem, [\Solomon], states that
\eqn\solo{
\prod_{i=1}^{d+1}(1+m_it)=\sum_{r=0}^{d+1} b_rt^r .
}
We have chosen the last exponent,
$m_{d+1}$, to be the one that equals unity. Therefore

$$
b_1=\sum^{d+1} m_i=\sum^dm_i+1,
$$

$$
b_2= \sum_{i<j}^dm_im_j+\sum^dm_i
$$
and
$$
b_3= \sum_{i<j<k}^{d+1}m_im_jm_k,\qquad b_4= \sum_{i<j<k<l}^{d+1}m_im_jm_km_l.
$$

Because the characteristic roots occur in complex conjugate pairs, we have
\eqn\numcirc{
b_1=\sum m_i=(d+1)h/2,
}
in terms of the Coxeter number (\cf \nocirc).
This result also follows from a generalisation of the equator argument
([\Coxeteraa] p 229-231, [\Steinbergb]). We also note that the largest exponent
equals $h-1$.

The breakdown
$$2g=\sum_rb_r$$
is obvious. Setting $t=-1$ in \solo\ we have
\eqn\beesum{
\sum_0^{d+1}(-1)^rb_r=0
}
and so
\eqn\really{
g=\sum_{\rm even} b_r=\sum_{\rm odd} b_r.
}
The first sum gives the order of the rotation subgroup, and the second the
(equal) number of remaining elements.

Returning to \identone, simple algebra yields
\eqn\fres{
{}_dS^{(d-1)}_1\big((d-1)/2\big)={1\over12g}\big(b_1(b_1-1)+b_2+(3-d)/2
\big).
}
(The systematic evaluation of the Bernoulli polynomials is discussed in the
appendix.)

When \fres\ is substituted into \identone, the last term cancels against
\firstt\ finally giving the identity
\eqn\identthreeb{
g\sum_{i<j}
\left({\pi\over\th_{ij}}-{\th_{ij}\over\pi}\right)|\pa{\cal M}_i\cap\pa
{\cal M}_j|=|\rS^{d-2}|\big(b_1(b_1-1)+b_2\big).
}

As a simple check we can set $d=2$, when $b_2=g-1$.
Using $|\rS^0|=2$, \identthreeb\ reduces to \identtwob. (As a minor point,
we note that
for $d=2$, the identity does not come from residues.)

For $d=3$, the easiest case to visualise, the $\pa{\cal M}_i$,
$(i=0,\,1,\,2,\,3,)$, are the spherical triangular faces of ${\cal M}$, the
spherical tetrahedron fundamental domain of the simplicial decomposition of the
3-honeycomb $\Pi_4,\,=\{k_1,\,k_2,\,k_3\}$.
The edges, $\pa{\cal M}_i\cap\pa{\cal M}_j$, are the six, circular sides of
these faces.
The situation is depicted in [\Coxetera] {\it Fig}. $\!7\scdot9$A,
[\Coxeterb] {\it Fig}. $\!4\scdot3$B, [\Goursat].
Verbal descriptions can be found in [\Coxeteraa] p 229, Todd
[\Todd] p 216, and Sommerville [\Sommerville] p 188, [\Sommervilleb].

The vertices $\P_i\,\,(i=0,\,1,\,2,\,3)$ of the fundamental domain
are obtained as
follows. $\P_3$ is the centre of a cell of the honeycomb, \ie of the projection
onto the circumsphere of a 3-face of the 4-polytope, $\Pi_4$. This cell
is itself a regular
(spherical) polytope, $\Pi_3$, and $\P_2$ is the centre of a cell of {\it its}
boundary. This cell is a 2-face (`face') of $\Pi_3$ (and of $\Pi_4$) and
is also a
regular polytope, $\Pi_2$. The cells, $\Pi_1$'s, of its boundary are 1-faces
(`edges') and $\P_1$ is
the centre of one of them. Finally, $\P_0$ is the end of such an edge and
is the one point that corresponds to a vertex of the original polytope,
$\Pi_4$.
An edge of a {\it cell} of the honeycomb has length $2|\P_0\P_1|$. A nested
set of $\Pi_i$ $(i=0,\,1,\,2,\,3)$ is called a {\it flag}. Cells are
sometimes referred to as {\it facets}.

Goursat [\Goursat] gives perhaps the most comprehensive description of the
geometry. He
refers to elements of the rotational subgroup, $\Ga^+$, as `transformations
droites' (\ie even) and to the remaining elements as `transformations gauches'
(\ie odd) and expresses them in terms of, up to four, inversions.

If a face of ${\cal M}$ be labelled dually by the opposing vertex,
then the dihedral angles, $\th_{ij}$, between faces $i$ and $j$ are given by
$\th_{01}=\pi/k_1$, $\th_{12}=\pi/k_2$, $\th_{23}=\pi/k_3$ with the rest
being
$\pi/2$'s ([\Todd], [\Coxetera], [\Sommervilleb]). $\th_{ij}$ can
also be characterised as the dihedral angle opposite to the edge $\P_i\P_j$.

We call the edge lengths $L_{ij},\,=|\P_{i}\P_{j}|$, and expand
the identity \identthreeb\ for $d=3$ as
$$
g\left(\big(k_1-{1\over k_1}\big)L_{23}+
\big(k_2-{1\over k_2}\big)L_{03}+
\big(k_3-{1\over k_3}\big)L_{01}+
{3\over2}\big(L_{12}+L_{02}+L_{13}\big)\right)
$$
\eqn\id{
\hspace{HHHHHHHHHHHHHHHH}=2\pi\big(b_1(b_1-1)+b_2\big).
}

To check \identthreeb, or \id, we need the edge lengths. Some of these
are given in [\Coxetera], p 139 and Table I (ii).

It is always the case that
$L_{01}+L_{23}+L_{03}$ and $L_{12}+L_{02}+L_{13}$ are separately
commensurable with $\pi$. For example, in $\{3^3\}$ these quantities are
both equal to $\pi$ (so that the total perimeter is $2\pi$) and
\id\ is then easily checked numerically using
$g=60$, $b_1=10$ and $b_2=35$.

General formulae for $L_{ij}$ exist
due to Sommerville [\Sommervilleb]. They are handy numerically and, for
convenience, are repeated here in the appendix. We have used them
to confirm $\id$ in all cases.

\sect{\bf 7. Geometrical meaning}

\noin A blind, numerical verification is all very well but equation
\identthreeb\
has a geometrical interpretation, or, if one prefers, a geometrical derivation.
The left-hand side is clearly a sum over all the fundamental domains of a
quantity associated with one domain and the right-hand side must be a global
measure of the same total quantity. We know, for example, that $b_1(b_1-1)/2$
is the number of $\rS^{d-2}$
intersections of $b_1$ $\rS^{d-1}$'s, including coincidences. The
corresponding identity will be derived from \identthreeb\ to see how the
various terms fit together. Our treatment is not the most direct.

For simplicity, attention is restricted to $d=3$.
It is convenient to depart from the traditional notation and define
\eqn\defkays{
k_{kl}={\pi\over\th_{ij}},\qquad(ijkl)\,{\rm cyclic}.
}

The two-dimensional case
suggests that we seek to evaluate the term
$$2g\sum_{k<l}{1\over2k_{kl}}L_{kl}$$
in \identthreeb.
Assembling $2g$ fundamental domains, the $1/2k_{kl}$ factor cancels
the multiplicity of appearance of the arc $\P_k\P_l$ and the result is the
total circumference of all the circular intersections of the reflecting
$\rS^2$'s. Hence
\eqn\essones{
2\pi b_2'=g\sum_{k<l}{1\over k_{kl}}L_{kl}
}
where $b_2'$ is the total number of complete $\rS^1$ boundaries in the
simplicial decomposition of the honeycomb. [For $\{3^3\}$, $b_2'$=25.]

The same reasoning applied to the perimeter gives the formula
\eqn\perima{
2g\sum_{k<l}L_{kl}=2\sum_{k<l}N^{kl}k_{kl}L_{kl}
}
where $N^{kl}$ is the number of $\P_k\P_l$ arcs in the simplicial
decomposition.  This agrees with

\eqn\edgenums{
N^{kl}={g\over k_{kl}}
}
which follows directly from the orbit-stabiliser theorem.
Using \essones, \identthreeb\ becomes (for $d=3$)
\eqn\ida{
g\sum_{k<l}k_{kl}L_{kl}=2\pi\big(b_1(b_1-1)+b_2+b_2'\big)\,.
}

Define, for convenience,
\eqn\perim{
D=g\sum_{k<l}L_{kl}-2\pi\big(b_2+b_2')
}
to get the intermediate result
\eqn\idb{
2\pi b_1(b_1-1)=
g\sum_{k<l}\big(k_{kl}-1\big)L_{kl}+D.
}
We will show that $D=0$, proving, incidentally, that the perimeter of the
fundamental domain is commensurate with $\pi$. We already have this because the
perimeter is half the sum of the four face-perimeters each of which is known
to be commensurate with $\pi$. We then see that $b_2+b_2'$ is the number of
$\rS^1$'s on all the bounding $\rS^2$'s, counting as distinct, circles that
belong to two, or more, $\rS^2$'s.

Proceeding in the geometrical vein, and noting that
the intersection degeneracy of the reflecting $\rS^2$'s at the
$(kl)$-edge is $k_{kl}\big(k_{kl}-1\big)/2$,
the total perimeter of the $\rS^1$ intersections, allowing for
coincidences, is
\eqn\idc{
2\pi b_1(b_1-1)=
\sum_{k<l}N^{kl}k_{kl}\big(k_{kl}-1\big)L_{kl}.
}
Hence, using \edgenums, a comparison with \idb\ shows that $D=0$, which
equality can be put into the form
\eqn\perimb{
2\pi b_2=\sum_{k<l}N^{kl}\,(k_{kl}-1)L_{kl}.
}

We have thus obtained identities, \idc\ and \perimb, precisely analogous to
\CFa\ and \trivident.

In $\oR^4$ the analogue of a rotation axis in $\oR^3$ is a rotation plane, \ie
an $\rS^1$ boundary on the circumsphere, $\rS^3$. Such a circle is divided
into a
number of arcs each with an associated order of rotation, one of the $k_{kl}$.
In words, \perimb\ says that $b_2$ is obtained by counting the number of
nontrivial rotations at every arc, weighted with that arc's relative size.

Although we can look upon our development as a {\it proof} of \perimb\ it ought
to be derivable directly. However, we have been unable, so far, to
provide a
geometrical reason for \perimb. It might be related to the various extensions
of the Euler formula such as the Dehn-Sommerville equations and other angle-sum
relations, of which there seem to be quite a variety, \eg [\Grunbaum].

To set the geometrical picture more closely, the general situation is again
recalled.

Each reflecting $(d-1)$-sphere is divided by its intersections with all the
others into a certain number of $(d-1)$-simplexes taken from those forming
the boundary of the fundamental domain. The topological structure is the
same for all the reflecting spheres although the composition may differ
metrically.

{}From \deemvol\ we find that the number of
$(d-1)$-simplexes in one reflecting $(d-1)$-sphere is $(d+1)g/b_1$. If we
assume \numcirc, this number equals $2g/ h$,
([\Coxetera], [\Coxeteraa] pp 231-232, [\Steinbergb]).

A simple example is the $\{3,4\}$ 2-honeycomb for which there are two types
of reflecting circles. One has the vertex content ${\bf 02120212}$ and the
other ${\bf 01010101}$. (See [\Coxetera] p 65.) Both have $2g/h=h+2=8$
segments. Incidentally this is a good example in which to check \ces.

We need to prescribe the polytope in more detail. This is done in terms of
its configurational numbers $N_{ij}$ ([\Coxetera] pp 12, 130, [\Sommerville],
[\Sommervilleb], [\Berger] \S$12\cdot6$). A regular polytope, $\Pi_n$ can be
thought of as a nested sequence of regular polytopes, of type
$\Pi_i$, ($i=0,\,1,\ldots,\,n$). (See [\Coxetera] pp 127-129.) An
element, $\Pi_j$, belongs to $N_{jk}$ $\Pi_k$'s for each
$k>j$ and contains $N_{jk}$ $\Pi_k$'s for each $k<j$. The number $N_i$,
which for symmetry's sake can be written $N_{ii}$, is just
the number of $\Pi_i$'s in the polytope. $N_0$ is the number of vertices, for
example, and $N_n$ is one.

For example, the regular simplex, $\al_{d+1}=\{3^d\}$, has a complete
symmetry group of order $(d+2)!$ and numbers, $N_i$,
\eqn\regsimp{
N_i= \comb{d+2}{i+1}.
}
(See [\Sommerville] p 96, [\Coxetera].)
Notationally, Coxeter in [\Coxetere] has $\big({}^{\textstyle i}|d+1\big)$ for
$N_i$.

If we set $n=d+1$ and take $\Pi_{d+1}$ to be the
Euclidean polytope that projects to the $d$-dimensional honeycomb, then the
vertices,
$\O_i$, ($i=0,\,1,\ldots,\,d$) of the $(d+1)$-simplex in the decomposition of
$\Pi_{d+1}$ project to the vertices $\P_i$ mentioned earlier (for $d=3$).
$\O_{d+1}$ is the centre of $\Pi_{d+1}$.

For the $(d+1)$-polytope $\Pi_{d+1}=\{k_1,k_2,\ldots k_d\}$ we can put
$$\Pi_i=\{k_1,\,k_2,\,\ldots,\,k_{i-1}\}$$
or, for the ascending construction,
$$\Pi'_i=\{k_{d-i+2},\,k_{d-i+3},\,\ldots,\,k_d\},$$
for the constituent polytope elements. The relation between the orders of the
complete symmetry groups of these $\Pi_i$'s and the numbers $N_i$ is
\eqn\enns{
N_i={|[k_1,\ldots,\,k_d]|\over|[k_1,\ldots,\,k_{i-1}]|\cdot
|[k_{i+2},\ldots,\,k_d]|}
}
which follows from the orbit-stabiliser theorem. The first factor in the
denominator is the order of the `internal' symmetry group of $\Pi_i$
while the second is the order of the stabiliser of $\Pi_i$ as a whole. The
ratio is then the index of the subgroup leaving the centre of $\Pi_i$, that is
$\P_i$, or $\O_i$, invariant \ie it is the number of $\Pi_i$'s. We must
remember that the internal symmetry group of an edge, $\{\}$, has order 2.
Equation \regsimp\ is an example of \enns. We note that $N_i=N'_{d-i}$.

Two special elements are $\Pi_d$, which is the `cell' of $\Pi_{d+1}$
(sometimes called the `bounding
figure'), and $\Pi'_d$, which is the `vertex figure'. A property of the regular
polytope is that the vertex figure of the bounding figure is the bounding
figure of the vertex figure. Continuing with the terminology, we may refer
to $\Pi_i'$ as the $(d+1-i)$-th vertex figure (\cf [\Coxetera] pp 133, 134.)
and the $\Pi_i$ as the $(d+1-i)$-th bounding figure. The $\Pi_i$ are the
{\it perischemons} of Schl\"afli.
\sect{\bf 8. Volume and degrees}

\noin A fundamental result of Schl\"afli's, [\Schl], concerns the variation
of the volume (or content) of a spherical polytope, $\Pi_d$, as the defining
parameters, such as the positions of the vertices, are altered slightly.
For arbitrary small displacements, the change in volume is
\eqn\var{
d|\Pi_d|={1\over d-1}\sum|\Pi_{d-2}|d\th
}
where the sum is over the facets, $\Pi_{d-2}$, of the various bounding
$\Pi_{d-1}$'s. (These facets
are Schl\"afli's {\it secondary perischemons}.) The angle $\th$ is that
between the $(d-1)$-planes intersecting in the facet. (In our previous notation
$\th_{ij}=\th$ and $\pa{\cal M}_i\cap\pa{\cal M}_j$ is a $\Pi_{d-2}$.) Equation
\var\ can be iterated on $d$.

We move quickly to $d=3$ and choose a 3-simplex for $\Pi_3$, \ie a
general spherical tetrahedron, for which the result was also given by
Richmond [\Richmond], [\Coxeterf]. Denoting the edges and corresponding
dihedral angles by $L$ and $\th$, the volume variation is
\eqn\voltet{
dV={1\over2}\sum_{\rm edges}Ld\th.
}
The tetrahedron is uniquely fixed by specifying the six $\th$'s or,
equivalently, the six $L$'s.

Schl\"afli points out the following `reciprocity'.
Consider another tetrahedron whose angles are $\pi-L$ and whose edges are
$\pi-\th$. The change in its volume is
\eqn\voltetr{
dV'=-{1\over2}\sum_{\rm edges}(\pi-\th)dL.
}
and so $d(V+V')$ is a perfect differential which integrates to
\eqn\recipr{
V+V'=\pi^2-{1\over2}\sum(\pi-\th)L.
}
The constant of integration can be determined by setting all the $\th$'s, and
$L$'s, equal to $\pi/2$ giving the self-reciprocal {\it orthant} of volume
$\pi^2/8$. Schl\"afli, [\Schlb] p 281, fixes the constant by choosing the sides
of $V$ vanishingly small, in which case $V'$ becomes {\it half} a full
sphere.

We wish to apply \recipr\ to a doubly rectangular tetrahedron (an
{\it orthoscheme}), such as a
fundamental domain, for which the angles are $\al$, $\be$, $\ga$ and
three $\pi/2$'s. Schl\"afli denotes the volume by $\pi^2f(\al,\be,\ga)/8$, so
normalising $f$ to unity for an orthant.

The reciprocal of a doubly rectangular tetrahedron
is a tetrahedron (not doubly rectangular in general) in which three
consecutive `nonplanar' edges equal $\pi/2$,
the rest being $\pi-\al$, $\pi-\be$ and $\pi-\ga$. We write its volume as
$\pi^2 F/8$.

Now we choose a fundamental domain, $\al=\pi/k_1$, $\be=\pi/k_2$,
$\ga=\pi/k_3$, and get from \recipr
\eqn\recipra{
f+F=8-{4\over\pi}\sum_{k<l}
\big(1-{1\over k_{kl}}\big)L_{kl},
}
where on the right we have reverted to our previous notation. Then, from
\perimb, using the fact that the volume of $2g$ fundamental domains
is $2\pi^2$, so that $gf=8$, we find
\eqn\reciprb{
b_2=g-1-{F\over f}.
}

We recall that, in the
two-dimensional (\ie $\oR^3$) case, $b_2=g-1$ on the simple geometric grounds
that rotations fix an axis but not a plane. In $\oR^4$, the elements
that fix a plane but not a hyperplane will certainly belong to the
rotation subgroup but now we have the extra possibility of a rotation
(a product of four reflections)
fixing a point but not a plane or hyperplane. These must also be removed,
as well as the identity. Hence we have
$$b_2=g-1-b_4,$$
which we already know, \really, and whence the curious relation
\eqn\curious{
F=b_4f\,.
}
The volume of the reciprocal tetrahedron is $b_4$ times that of the
fundamental orthoscheme. Numerically, $b_4$ is the product of the exponents,
$m_j$. For $\{3^3\}$, $f=2/15$ and $b_4=24$. At the moment, we have no direct
geometrical interpretation of \curious. The reciprocal orthoscheme does not
seem to be a particularly useful construct.


Schl\"afli [\Schlb] interestingly uses \recipr, applied to a polytope, to
deduce several numerical results for $f$.

We remark that the Schl\"afli function, $f$, has been used
occasionally in physics, [\Leese], [\Laursen], [\Woit].
\vfill\eject
\sect{\bf 9. A further identity}

\noin A further identity follows on setting $m=3$ in \hkcoeff,
$$
C_{3/2}={8\pi^{(d+1)/2}\over\Ga\big(d/2-1\big)}{}_dS^{(d-2)}_1$$

\eqn\identthreec{
={\pi^{3/2}\over 3g}|\rS^{d-3}|\, B_3^{(d)}\big((d-1)/2\mid{\bf d}\big)
={\pi^{3/2}\over 24g}|\rS^{d-3}|\,(2b_1(b_1-b_2+d-3)),
}
(see the appendix). This will come into play for $d\ge 3$.

We can only state the result for the (Neumann) coefficient $C_{3/2}$ in the
incomplete form,
\eqn\ceethreehalves{
C_{3/2}={\sqrt\pi\over192}\sum_i\int_{\pa{\cal M}_i}\big(6\tr(\ka^2)-3\ka^2
-4\wR+12(8\xi-1)R\big)dS+{\rm corner}\,\,{\rm terms}
}
where $\wR$ is the intrinsic curvature of the boundary parts.

In $d$-dimensions, the curvature part of $C_{3/2}$ reduces to
\eqn\curv{
{\sqrt\pi\over24}(d-1)(d-2)|\pa{\cal M}|.
}
This combines with the last term in \identthreec, if we use \Lapd\ for
$b_1$, to give for the (unknown) corner terms, the rather simple value
\eqn\corner{
{\rm corner\,\,terms}={\pi^{3/2}\over 12g}|\rS^{d-3}|\,b_1
\big(b_1-b_2-2\big) <0,
}
in this particular geometry. We check that the right-hand side vanishes for
$d<3$.

To the author's knowledge, no {\it general} form for the corner term
is known. An expression in the case that ${\cal M}$ is a polygonal cylinder
is given by Pathria [\Pathria] and Baltes, [\Baltes], but this is not enough.
There are some general results of Cheeger [\Cheeger]. For the time being, we
can just take the attitude that we are {\it calculating} these corner terms
for a special geometry.

The derivation of \ceeone\ can be traced back to work of Sommerfeld on
propagation in a wedge of arbitrary angle. To get the missing terms in
\ceethreehalves\ it would be necessary to discuss a general trihedral corner.

It would be interesting to use the result \corner\ to get a handle on the
form of the corner terms. If $d=3$, the right-hand side of \corner\ must
have a combinatorial-geometric significance in terms of the vertices of the
simplicial decomposition of $\rS^3$. In terms of the order $g$, we find the
value of $b_1(b_2-b_1+2)$ to be $9g/2$ for $\{3^3\}$, $6g$ for $\{3^24\}$, $7g$
for $\{343\}$ and $9g$ for $\{3^25\}$. For the record, the corresponding
values of $b_2-b_1+2$ are 27, 72, 168 and 1080.

Let us recapitulate the identities. So far we have \emvol, \deemvol,
\identthreeb\ and \identthreec\ (with \ceethreehalves). Assume that the
volume of ${\cal M}$, and in fact all the geometry of
${\cal M}$, is known in terms the dihedral angles. Then, with
Schl\"afli, the order, $2g$, of the symmetry group can be found from \emvol.
Next, $b_1$ follows from \deemvol, and $b_2$ from \identthreeb\
but $b_3$ is curiously absent from \identthreec.



\sect{\bf 11. Higher relations}

\noin We return to two dimensions and continue with the discussion in section
4.

More generally than looking at $\ze(0)$,
we have from the equality of $g\ze(-k)$,
\eqn\geneq{
\sum_mmn_m\ze_m(-k)-2\big(\sum_mn_m-1\big)\ze_m\big(-2k-1,{1\over2}\big)=
{2\over(2+k)(1+k)}B^{(2)}_{2+k}\big({1\over2}\mid d_1d_2\big).
}
The Bernoulli functions are `non-topological'.

We repeat the expression for $\ze_m(-k)$ given in [\Chang]
\eqn\ber{
\zeta_m(-k)=-{1\over m(k+1)}B_{2k+2}(1/2)+
{m^{2k-1}\over2k+1}\sum_{p=0}^{m-1}(2p+1)B_{2k+1}\!\big((2p+1)/2m\big).
}

The following identity, derived in the appendix to [\Chang], is needed in
order to rewrite \ber\ in a form more suitable for substituting into
\geneq.

\eqn\equiva{
\sum_{l=0}^k {\comb{2k}{2l}} \,(m^{2l}-1)
B_{2k-2l}\big({1\over2}\big)\, B_{2l}
=
k\,m^{2k-2} \sum_{p=0}^{m-1}
(2p+1)\, B_{2k-1}\big((2p+1)/2m\big) \,.
}

For \geneq\ we then find
$${2\over(2+k)}B^{(2)}_{2+k}\big({1\over2}\mid d_1d_2\big)=$$
\eqn\geneqb{
{1\over2k+1}\sum_mn_m\sum_{l=1}^{k+1}\comb{2k+2}{2l}\big(m^{2l}-1\big)
B_{2l}B_{2k+2-2l}\big({1\over2}\big)-B_{2k+2}\big({1\over2}\big).
}
For $k=0$ this is our previous result of course. We write out the explicit
form when $k=1$,
\eqn\kone{
240B^{(2)}_3\big({1\over2}\mid d_1d_2\big)=-4\sum_mn_m\big(2(m^4-1)
+5(m^2-1)\big)-21<0.
}

Generally, looking at the form of \geneqb, we see that by a recursion
argument
we can derive an expression for the sum $\sum_ln_m(m^{2l}-1)$ in terms of
the degrees. Then, making use of the relations \reg, and of \euler, this
expression will yield an equation for the sums of all odd powers of the
three orders, $\sum_l m^{2l-1}$, in terms of the two degrees.

There is no doubt that these relations are numerically satisfied. After
all they are identities. They have nevertheless been checked for the regular
solids. (The dihedral case $[m]$ may also be used as an example. For this,
$(p,q,r)=(m,m,1)$ and $d_1=m,\,d_2=1$. Heuristically we can set
$n_p=n_m=1/2$ rather than make this a special case.)

$B^{(2)}_3\big({1\over2}\mid d_1d_2\big)$ can be calculated or looked up.
It comes out directly as
$$
240B^{(2)}_3\big({1\over2}\mid d_1d_2\big)=
-8(d_1^4+d_2^4)+40d_1^2d_2^2-120d_1d_2(d_1+d_2)+$$

\eqn\beetwo{
60(d_1^2+d_2^2+3d_1d_2)-60(d_1+d_2)+15
}
but it should be rewritten in terms of $g$ and $b_1$. We have the easily
verified relation
$$
d_1^4+d_2^4=b_1^4+4b_1^3-4gb_1^2+6b_1^2+2g^2-8gb_1-4g+4b_1+1
$$
and we find for the left-hand side of \kone,
\eqn\konelhs{
-8b_1^4-32b_1^3+12b_1^2+28b_1+32gb_1^2-56gb_1+24g^2-28g+7.
}

The resulting identity cannot be obtained from the C-F relation. So far as
I can see it does not have a geometrical interpretation. Since there are only
a well defined set of orders $(p,q,r)$ it is perhaps
not surprising that there are many particular relations.

\sect{\bf 12. Conclusion}

\noin Since the approach depends on the expansion of the heat-kernel, it is
restricted by the availability of explicit forms for the coefficients. It is
certainly of interest to obtain the unknown corner terms.

Because of our inability to derive \perimb\ directly (it may of course be
obvious), we would like to claim a modest amount of novelty for our method.

At one level, it might be said that all we are doing is checking
known coefficients in a specific situation. There is, however, a little more
to it than that. To check the numbers, it is not necessary to rewrite
the right-hand side of \ceeone\ in terms of the $b_r$. It could have been left
in the degrees. However the dramatic simplification of the Bernoulli functions
and the various cancellations suggest some interesting general relations that
have yet to be uncovered.

If it is wished to check numerically the identity for $d=4$, say, then it
will be necessary to work out the {\it areas} of the 2-faces between
neighbouring 3-face cells of the boundary of the 4-simplex fundamental domain.

It does not seem possible to extend easily the methods of this paper to
the stellated
polytopes, such as the Kepler-Poinsot polyhedra. The difficulty is the
evaluation of the heat-kernels on the branched coverings. In such highly
symmetrical situations it might be possible to obtain closed forms for the
heat-kernels  but we have not been able to make any progress in this
direction. Laporte [\Laporte] has some useful comments on the mode
problem.

In the two-dimensional
case, Rubinowicz [\Rub] claims to have a method of integrating the
wave-equation on an
arbitrary Riemann surface, but, in fact, only discusses the case of one
branch point in detail and says that the general case follows upon a process of
`Zusammenst\"ucklung', which, unfortunately, he does not exhibit.

\sect{\bf Appendix}

\noin We give here Sommerville's expressions [\Sommervilleb] for the edge
lengths of
the fundamental simplex of the $d$-dimensional spherical honeycombs.
For  $\{3,3,\ldots,w\}$ $=\{3^{d-1},w\}$,

\eqn\lengthsa{
\sin^2L_{ij}={j-i\over j+1}\cdot{1-d\cos\big(2\pi/w\big)
\over1-(d-i-1)\cos\big(2\pi/w\big)}
}
while, for the reciprocal, $\{w,\ldots,3,3\}=\{w,3^{d-1}\}$,
\eqn\lengthsb{
\sin^2L_{ij}={j-i\over d-i+1}\cdot{1-d\cos\big(2\pi/w\big)
\over1-(j-1)\cos\big(2\pi/w
\big)}.
}
For the more general case, $\{v,3,3,\ldots,3,3,w\}=\{v,3^{d-2},w\}$,
\eqn\lengthsc{
\cos^2L_{ij}={1+(1+j-d)\cos\big(2\pi/w\big)\over1+(1+i-d)\cos
\big(2\pi/w\big)}
\cdot{1+(1-i)\cos\big(2\pi/v\big)\over1+(1-j)\cos\big(2\pi/v\big)}.
}
Finally for the case $\{3,4,3,3,\ldots,3\}=\{3,4,3^{d-2}\}$ we have
$$
\cos^2L_{ij}={d-j+1\over d-i+1}\cdot{5-i\over5-j}
\qquad(1\le i<j\le d)
$$
\eqn\lengthsd{
\cos^2L_{0j}={d-j+1\over5-j}
}
and for its reciprocal, $\{3,3,\ldots3,4,3\}=\{3^{d-2},4,3\}$,
$$
\cos^2L_{ij}={5+j-d\over5+i-d}\cdot{i+1\over j+1}
\qquad(i<j\le d-1)
$$
\eqn\lengthse{
\cos^2L_{i d}={i+1\over5+i-d}.
}

We now turn to the connection of the generalised Bernoulli numbers with the
Todd Polynomials.

Hirzebruch [\Hirz] gives the relation
\eqn\Toddp{
T_k(c_1,\ldots,c_k)={(-1)^k\over k!}B^{(d)}_k(d_1,\ldots,d_d)\quad k\le d
}
where, the $c_s$ are the elementary symmetric functions of the degrees $d_i$
($i=1,\ldots, d$).

Using the known formulae for the $T_k$ we can obtain
the Bernoulli quantities if some transformations are made. We wish to
write the quantities in terms of the elementary symmetric functions,
$b_r$, of the {\it exponents} $m_i$ ($i=1,\ldots,d+1$) so we firstly
need the connection between the $c_s$ and the $b_r$. This is easily found
as follows. First divide the polynomial,
$$\prod_1^d(1+m_it)={1\over1+t}\sum_0^{d+1}b_rt^r\equiv\sum_0^d b_r't^r
$$where
$$b_r'=(-1)^{r+1}\sum_{r+1}^{d+1}(-1)^i\, b_i=(-1)^r\sum_0^r(-1)^i\,b_i$$
using \beesum. Then set $t\rightarrow t/(1+t)$ to get
$$
\sum_0^db_r't^r(1+t)^{d-r}=\prod_1^d(1+d_it)=\sum_0^d c_st^s.
$$
Expanding the $(1+t)$, the $c_s$ can be read off and a rearrangement of
the summations gives
\eqn\cees{
c_s=\sum_{r=0}^s(-1)^rb_r\sum_{i=r}^s(-1)^i\comb{d-i}{d-s}.
}

As a point of more than technical interest, Lemmas 1 and 2 of Todd [\Toddb]
provide an alternative means of obtaining the elementary symmetric functions of
quantities translated by a constant.

More generally than is needed here we have the following result
$$
c_s\big(\{d_i+\la\}\big)=\sum_{i=0}^s\la^i\comb{d+i-s}{i}c_{s-i}
\big(\{d_i\}\big)
$$
which follows directly upon the replacement $t\rightarrow t/(1+\la t)$ in
$$\prod_1^d(1+d_it)=\sum_o^dc_s\big(\{d_i\}\big)t^s$$
or from a Taylor expansion and then use of the standard formula
$$
{\pa c_s\big(\{d_i\}\big)\over\pa d_{i}}=sc_s\big(\{\hat d_i\}\big)
$$
where the hat signifies omission of the term,
(\cf Lemma 2 in [\Toddb]). Note that Todd's $h_s$ is a sum of {\it all}
homogeneous products \ie
$$\sum_0^dh_st^s={1\over \sum_0^d(-1)^sc_st^s}={1\over\prod(1-d_it)}.
$$

The quantity we actually require is the Bernoulli function
$B^{(d)}_r\big(x|{\bf d}\big)$ (at $x=(d-1)/2$) which is a polynomial in
$x$ whose coefficients are the Bernoulli numbers in \Toddp,
\eqn\bernf{
B^{(d)}_r\big(x\mid{\bf d}\big)=(-1)^rr!\sum_{l=0}^r(-1)^l{x^l\over l!}
T_{r-l}\big(c_1,\ldots,c_{r-l}\big).
}

{}From the listed expressions for the Todd polynomials in the $c_s$, using
\cees\ and \bernf, we can generate the Bernoulli polynomials at $x=(d-1)/2$
in terms of the
$b_r$. These have been used in the main text. We collect a few results here.
$$B^{(d)}_2\big((d-1)/2\mid{\bf d}\big)={1\over6}\big(b_1(b_1-1)
+b_2+(3-d)/2\big)
$$

\eqn\genbern{
B^{(d)}_3\big((d-1)/2\mid{\bf d}\big)={1\over4}b_1\big(b_1-b_2+(d-3)/2\big)
}
$$B^{(d)}_4\big((d-1)/2\mid{\bf d}\big)=-{1\over240}\big(8b_4-24b_3-8b_3b_1
-24b_2^2-4b_2(8b_1^2-16b_1-5d+9)+$$
$$\hspace{MMMMMMMM}
4b_1(b_1-1)(2b_1^2+10b_1+5d-13)-(5d-3)(d-5)\big).$$

It is probably more instructive to evaluate the Bernoulli polynomials
directly rather than go through the Todd polynomials. We can use the general
theory of multiplicative series as developed by Hirzebruch [\Hirz] and this
will be detailed elsewhere.
 
  \vfill\eject\immediate\closeout\reffile
  \noindent{{\bf References}}\bigskip\frenchspacing

  \input refs.tmp\vfill\eject\nonfrenchspacing
\end